\documentclass[aps,prb,twocolumn,superscriptaddress,showpacs]{revtex4-1}
\usepackage{dcolumn}
\usepackage{bm}
\usepackage{amssymb}
\usepackage{amsmath}
\usepackage{amsfonts}
\usepackage[english]{babel}
\usepackage{amsmath}
\usepackage{graphicx}
\usepackage{dcolumn}
\usepackage{bm}
\usepackage{hyperref}

\usepackage[usenames, dvipsnames]{color}

\usepackage[]{graphicx}

\usepackage{epstopdf}

\begin{document}

\preprint{AIP/123-QED}

\title[]{Study of the velocity plateau of Dzyaloshinskii domain walls}
\thanks{\it{This paper is dedicated to the memory of John C. Slonczewski.}}

\author{V. Krizakova}
\thanks{Present address : Laboratory for Magnetism and Interface Physics, Department of Materials, 
ETH Zurich, 8093 Zurich, Switzerland}
\affiliation{Univ. Grenoble Alpes, CNRS, Institut N\'{e}el, F-38000 Grenoble, France}
\author{J. Pe\~{n}a Garcia }
\affiliation{Univ. Grenoble Alpes, CNRS, Institut N\'{e}el, F-38000 Grenoble, France}
\author{J. Vogel}
\affiliation{Univ. Grenoble Alpes, CNRS, Institut N\'{e}el, F-38000 Grenoble, France}
\author{N. Rougemaille}
\affiliation{Univ. Grenoble Alpes, CNRS, Institut N\'{e}el, F-38000 Grenoble, France}
\author{D. de Souza Chaves}
\affiliation{Univ. Grenoble Alpes, CNRS, Institut N\'{e}el, F-38000 Grenoble, France}
\author{S. Pizzini}
\email[]{stefania.pizzini@neel.cnrs.fr}
\affiliation{Univ. Grenoble Alpes, CNRS, Institut N\'{e}el, F-38000 Grenoble, France}
\author{A. Thiaville}
\email[]{andre.thiaville@u-psud.fr}
\affiliation{Laboratoire de Physique des Solides, Univ. Paris-Sud, CNRS UMR 8502, 91405 Orsay, France}


\date{\today}

\setlength\abovecaptionskip{2pt}
\setlength\belowcaptionskip{-2pt}

\begin{abstract}
We study field-driven domain wall (DW) velocities in asymmetric multilayer stacks with perpendicular magnetic
anisotropy and Dzyaloshinskii-Moriya interaction (DMI), both experimentally and by micromagnetic simulations.
Using magneto-optical Kerr microscopy under intense and nanoseconds-long fields, we show that DWs in these films
propagate at velocities up to hundreds of m/s and that, instead of the expected decrease of velocity after the
Walker field, a long plateau with constant velocity is observed, before breakdown.
Both the maximum speed and the field extent of the velocity plateau strongly depend on the values of the spontaneous
magnetization and the DMI strength, as well as on the magnetic anisotropy.
Micromagnetic simulations reproduce these features in sufficiently wide strips, even for perfect samples.
A physical model explaining the microscopic origin of the velocity plateau is proposed.

\end{abstract}


\maketitle

\section{\label{sec:intro} Introduction}

The interfacial Dzyaloshinskii-Moriya interaction (DMI) \cite{Dzyaloshinskii1957,Moriya1960} in ultra-thin
magnetic films with
perpendicular magnetic anisotropy (PMA) has undergone intensive studies in recent years, due to the peculiar field-
and current-driven dynamics of non-collinear magnetic textures such as chiral domain walls (DWs) or magnetic skyrmions
which this interaction can stabilize \cite{Heinze2011, Chen2013, Fert2013}.
The most important attribute of the DWs gained by the effect of this DMI is their fixed chirality together with the
N\'{e}el internal structure \cite{Thiaville2012}.
The first consequence of this is the shift of the Walker field to larger values, enabling the DWs to reach higher
velocity with respect to systems with vanishing DMI \citep{Thiaville2012}.
These DWs can also be moved efficiently by an electrical current via the spin-orbit torque associated to the spin Hall
effect (SHE-SOT), reaching higher velocities compared to most systems without DMI
\cite{Moore2008,Miron2011,Ryu2013,Emori2013}.
Successive DWs move along the same direction as required for race-track applications \cite{Parkin2008b},
making these so-called Dzyaloshinskii DWs attractive for applications in spintronics.

As is well known, one-dimensional (1D) micromagnetics predicts that the velocity of field-driven DWs increases
with the applied magnetic field, in a steady-state flow regime until reaching a maximum at a certain
threshold, called the Walker field ($H_\text{W}$) \cite{Walker1956,Slonczewski1972,Schryer1974,Malozemoff1979}.
Above it, a sharp decrease of the velocity - called Walker breakdown - occurs, down to the regime of
precessional motion where the DW magnetization continuously rotates \cite{Slonczewski1972,Malozemoff1979},
with a constant and lower mobility (the ratio of velocity to field).
As remarked early \cite{Slonczewski1972}, the breakdown region in which the DW differential mobility is negative,
is anticipated to be unstable when the sample is two- or three-dimensional.
Indeed, any deformation with respect to a straight DW is, in the negative mobility regime, amplified.
Experimentally, a plateau of velocity beyond the Walker field has often been reported.
For bubble garnet films, it was called limiting \cite{Bobeck1971}, asymptote \cite{Argyle1972} or saturation velocity
\cite{deLeeuw1978,Malozemoff1979,Volkov1982}.
A simple 2D model based on the periodic creation and annihilation of horizontal Bloch lines \cite{Slonczewski1972b} moving
along the sample thickness provided a first analytic expression for this velocity.
To describe experimental results, empirical formulae for this velocity were also proposed \cite{deLeeuw1978,Volkov1982}, but
without physical justification.
More recent experiments have also reported the phenomenon, sometimes more complex than just a plateau.
For example, for (Ga,Mn)As films with perpendicular magnetization, a plateau and velocity oscillations were
observed \cite{Dourlat2008}.
These phenomena could be physically explained by the coupling of DW global motion to its flexural modes across
the thickness, the latter involving both DW position and magnetization \cite{Gourdon2013}.
Thus it appears that, generally, the DW dynamics above the Walker field is controlled by processes within the wall,
the nature of which depends on the sample characteristics.

In ultrathin films with PMA, the presence of interfacial DMI also strongly modifies the DW dynamics above the Walker field,
as shown experimentally \citep{Yoshimura2015,Pham2016,Ajejas2017}, and reproduced by
micromagnetic large-scale simulations \citep{Yoshimura2015,Yamada2015,Pham2016}.
For sufficiently large DMI, the abrupt DW velocity decrease after the Walker field, expected from 1D theory, is replaced
by a velocity plateau, before the DW enters the precessional regime for larger fields.
This holds provided the DW length (the width of the sample when patterned in a strip) is sufficient;
for narrower strips, up to two breakdowns were observed before reaching the plateau \cite{Yamada2015}.
As ultrathin samples are of a thickness well below the micromagnetic exchange length, neither the creation and motion of
horizontal Bloch lines seen in bubble garnet films, nor the excitation of perpendicular standing spin waves in the
DW \cite{Gourdon2013} can be invoked to explain the phenomenon.
On the other hand, micromagnetic simulations highlight the presence of vertical Bloch lines (VBLs) beyond the Walker field.
Their creation and annihilation has been claimed to explain the presence of the velocity plateau \cite{Yoshimura2015}.
In this paper, motivated by experimental observations, we have studied this phenomenon in detail by analysing the
DW structure, the VBLs, and their dynamics for samples presenting different magnetic properties.
This allows us to explain the correlation of the velocity plateau with the number of $2 \pi$ VBLs, and quite surprisingly
to link the field marking the end of the plateau with the DMI field stabilizing the DWs in the N{\'{e}}el structure.

\begin{table*}[t]
\caption{Measured magnetic parameters for the samples whose velocity curves are shown in 
Figure~\ref{fig:Fig1-speeds}, and comparison of the measured
maximum velocity $v_\text{max}$ and end of velocity plateau $B_\text{break}$ with simulation results
using these parameters, together with $\alpha=0.15$, exchange being $A_\text{ex}=16$~pJ/m for (i) and (ii) 
and $A_\text{ex}=4$~pJ/m for (iii) and (iv).
The last column shows the Slonczewski field $B_\text{S}$ estimated with Eq.~(\ref{eq:H_S_DDW}).}
\label{tab:param}
\begin{center}

\begin{tabular}{ c  c c c c  c c c c c  }
\hline \hline \\
sample &  $M_\text{s}$ &  $D$  & $K_\text{eff}$ & $D/M_\text{s}$  &  $v_\mathrm{max}^\mathrm{exp}$  &
$v_\mathrm{max}^\mathrm{sim}$ & $B_\mathrm{break}^\mathrm{exp}$ &  $B_\mathrm{break}^\mathrm{sim}$
&  $B_\text{S}$  \\  \hline \\
           & 10$^{5}$~A/m  & mJ/m$^2$ & MJ/m$^{3}$ & nJ/(A.m) & m/s & m/s & mT & mT & mT  \\\hline \\\\

(i)~Pt/Co(1~nm)/Gd     & 6.4  & 1.45  & 0.26 & 2.3  &  640 &  600 & $>$275  & 265 & 325 \\\\

(ii)~Pt/Co(1~nm)/GdOx   & 12.6 & 1.5   & 0.44 & 1.2  & 300  & 280 & $>$175 &  200 & 225 \\\\

(iii)~Pt/GdCo(4~nm)/Ta  & 2.3  & 0.2  & 0.08 & 0.9  & 250  & 230 &   130  &  125 & 140 \\\\

(iv)~Pt/GdCo(4.8~nm)/Ta  & 3.5 & 0.2  & 0.06 & 0.6  & 160  & 140  &   60 &   35  & 80  \\\\

\hline
\hline
\end{tabular}
\end{center}
\end{table*}

\section{\label{sec:exp} Experiments}

\begin{figure}
\includegraphics[width=1\columnwidth]{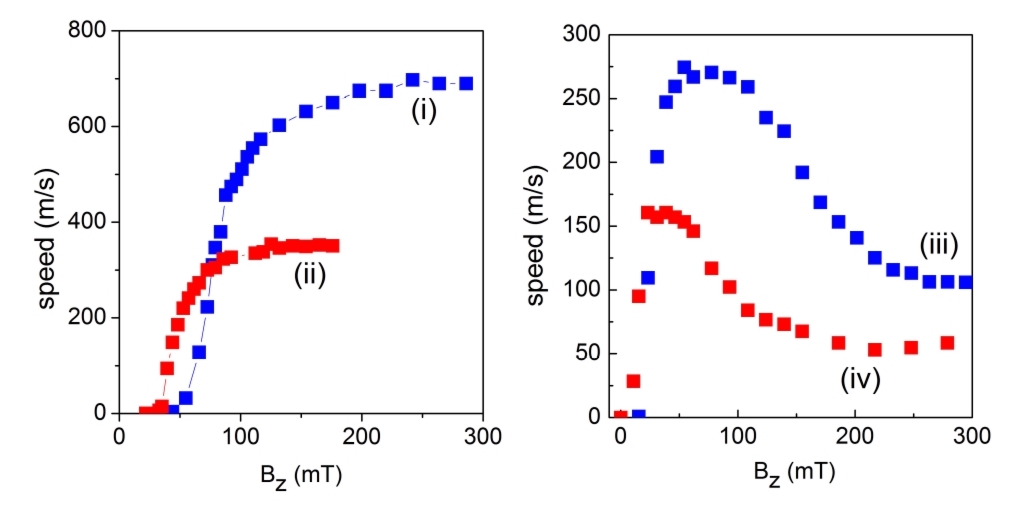}
\caption{\label{fig:Fig1-speeds}
Domain wall velocity versus easy-axis field $B_z$ measured for (i) Pt/Co(1~nm)/Gd, (ii) Pt/Co(1~nm)/GdOx,
(iii) Pt/GdCo(4~nm)/Ta and (iv) Pt/GdCo(4.8~nm)/Ta.
}
\end{figure}

Domain wall velocity versus easy axis magnetic fields $B_{z}$ were measured for four samples characterized by very
different values of spontaneous magnetization $M_\text{s}$ and DMI strength $D$, which, as we will show, are at the
origin of the different behaviours of the domain wall dynamics: (i) Pt/Co(1~nm)/Gd/Al; (ii)Pt/Co(1~nm)/GdOx/Al;
(iii)~Pt/GdCo(4~nm)/Ta and (iv)~Pt/GdCo(4.8~nm)/Ta (layers are listed from bottom to top).
The first two multilayer stacks were studied in details in Ref.~\onlinecite{Pham2016}.
The samples were prepared by magnetron sputtering on Si/SiO$_{2}$ substrates \citep{Pham2016}.
The Gd$_{x}$Co$_{1-x}$ layers with composition gradient $x~\sim{0.21-0.23}$ were prepared by co-sputtering of
Gd and Co targets \citep{Chaves2018}.
Because of the vicinity of the composition compensation at room temperature (RT), the magnetization is strongly
reduced in these two samples.
Moreover, since the compensation temperature in these alloys is very sensitive to the composition, their RT
magnetizations differ.
The material parameters of the four samples are summarized in Tab.~\ref{tab:param}.
Magnetization $M_\text{s}$ and effective anisotropy $K_\text{eff}$ were measured by magnetometry, whereas
the DMI parameter $D$ was estimated from the in-plane field dependence of the DW velocity, in the flow regime.
The details of the Kerr microscopy experiments allowing the measurement of DW velocities as a function of
B$_{z}$, and of the DMI strengths are also described in Ref.~\onlinecite{Pham2016}.
In that work, we showed that the field-driven domain wall velocity after the Walker field is tuned by the ratio
between the DMI strength [$D\sim{1.5}$~mJ/m$^{2}$ for samples (i) and (ii)] and the spontaneous magnetization
$M_\text{s}$.
While for Pt/Co/GdOx, $M_\text{s}$ at RT is close to the bulk value ($M_\text{s}$=1.26 MA/m), a strong reduction
of ($M_\text{s}$=0.64 MA/m) is observed in Pt/Co/Gd, where an interfacial ferrimagnetic alloy forms
at the top Co interface.
The saturation DW velocity at high field, of the order of 300~m/s in Pt/Co/GdOx, increases up to 600~m/s in
Pt/Co/Gd, due to the lower magnetization.
In both samples, the velocity plateau extends up to the  largest fields for which the speeds can be measured
[Figure~\ref{fig:Fig1-speeds}(a)].
The situation is different for the Pt/GdCo/Ta trilayers where the magnetization $M_\text{s}$ as well as the
DMI are reduced (see Tab.~\ref{tab:param}).
The maximum velocities are smaller than in the previous samples and moreover, while in  sample (iii) a drop of the
velocity is observed around 230~mT, in sample (iv) the velocity drops soon after the Walker field, at around 60~mT.
The saturation velocity and the velocity breakdown field for the four samples are reported in
Tab.~\ref{tab:param}.
In the following, we show that 2D numerical simulations reproduce quantitatively the different behaviour of
the DW velocity in the four samples, and propose a physical understanding of all the results.

\section{\label{sec:method} Calculation methods}

\begin{figure}[b]
\includegraphics[width=1\columnwidth]{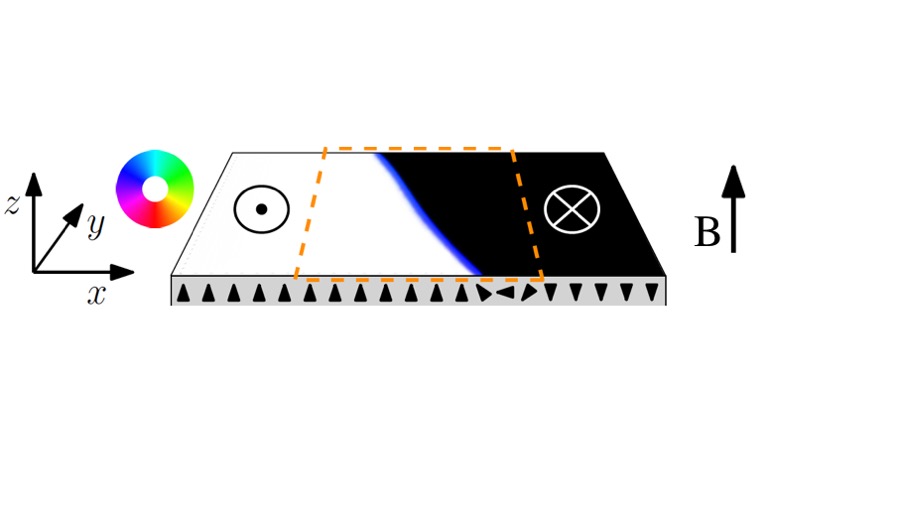}
\caption{\label{fig:scheme} Scheme of simulation geometry with depicted micromagnetic spin configuration.
White (black) corresponds to magnetization upward (downward), and the color wheel shows the orientation of
the in-plane magnetization component.
The dashed square depicts the moving calculation window.}
\end{figure}

Micromagnetic simulations were realized using the \mbox{Mumax3} software \citep{Vansteenkiste2014} which solves
the Landau-Lifshitz-Gilbert equation in finite-differences discretization.
The initial material parameters $M_\text{s} = 1.26$~MA/m, \mbox{$D = 1.5$~mJ/m$^2$,} $K_\text{eff} = 0.44$~MJ/m$^3$,
$A_\text{ex} = 16$~pJ/m and magnetic damping constant $\alpha = 0.15$ were selected to imitate the magnetic
properties of sample (ii).
Although disorder is absent in the simulations presented here, we have verified that it does not affect the main
results of this work.
Subsequently, $M_\text{s}$, $K_\text{eff}$ and $D$ were varied in order to cover the range of values obtained experimentally
for the other trilayers.
In the simulations, a N{\'{e}}el wall introduced into the strip-shaped sample is displaced by the action of a magnetic
field normal to the plane, as sketched in Fig.~\ref{fig:scheme}, within a~$(1 \times 1)~\mu\text{m}^2$ moving-frame window
(so as to keep the domain wall in its center).
The lateral mesh is chosen to be $\approx (2 \times 2)~\text{nm}^2$ as it is sufficiently accurate with respect to
the DW width parameter $\Delta = \sqrt{A_\text{ex}/K_\text{eff}}$, equal to 5-10~nm in the cases examined here.

The dynamics of a DW in perpendicularly magnetized nanostrips of different widths was first studied, to explore
the transition between the 1D and the 2D behavior.
For widths larger than 500~nm the DW dynamics displayed the characteristics observed experimentally, i.e. large DW
speeds with a plateau of maximum velocity.
Consequently, the strip width of 1~$\mu$m was selected for all the following computations.

The micromagnetic configurations were analyzed by first finding the DW precise path.
Having located the DW at the bottom edge of the calculation box, the DW was followed
as it crossed successive horizontal or vertical mesh lines between the mesh points.
To analyze the DW magnetic structure, the tangential and normal components of the DW (in-plane) magnetization
were then computed.
In addition, the local in-plane angle $\varphi$ of the DW magnetization, counted from the DW tangent and,
for continuity, not restricted to the $[0, 2 \pi]$ interval,  was evaluated starting from the strip $y$-bottom
(see axes definition in Fig.~\ref{fig:scheme}).
Thus, $\varphi=0$ is a left-handed Bloch Wall, $\varphi= \pi/2$ is a left-handed N{\'{e}}el wall, etc.
VBLs were identified by processing the profile of $\varphi$ versus curvilinear abscissa $s$ along the DW.
The statistics shown below (number and density of $2 \pi$ VBLs, DW length, ..) were obtained on 500 snapshots from
10 DW configurations, obtained by running a single calculation for 500~ns and collecting 50 snapshots at 1~ps
interval, with a 50~ns interval between two configurations.
Indeed, as the DW configuration does not change enough on the ns time scale, measurements on a
single DW configuration show excessive scatter.

\section{\label{sec:solitons} Simulation results and analysis}

\begin{figure}
\includegraphics[width=0.85\columnwidth]{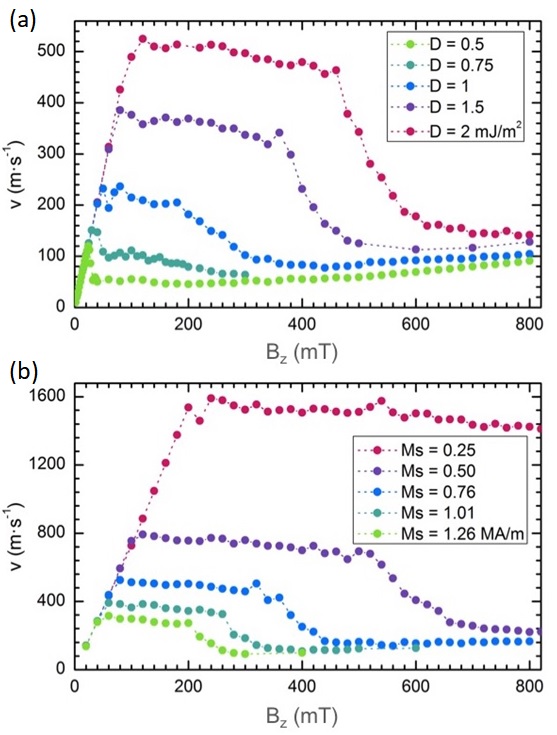}
\caption{\label{fig:speed-Ms-DMI}
Simulated field-driven DW velocity for (a) $M_\text{s}$=1.01~MA/m and $K_\text{u}$=1.44~MJ/m$^3$ so that
$K_\text{eff}$=0.80~MJ/m$^3$, and different values of DMI strength and
(b) fixed $D$=1.5~mJ/m$^2$, $K_\text{eff}$=0.44~MJ/m$^3$ and varying spontaneous magnetization.
The different lengths of the velocity plateau for $D$=1.5~mJ/m$^2$ and
$M_\text{s}$=1.01~MA/m in (a) and (b) result from the different $K_\text{eff}$ values (see discussion
later in the text).
}
\end{figure}

Figures~\ref{fig:speed-Ms-DMI}(a,b) show the simulated velocity vs. field curves for different $D$ or
$M_\text{s}$ values while keeping the other parameters constant.
The DW velocity $v_\text{W}$ at the Walker field $H_\text{W}$ is seen to increase linearly with the DMI
strength $D$, and to decrease as the magnetization $M_\text{s}$ increases.
These results agree with the previously derived analytic formula (the last equality
holding when the DMI field is much larger than the internal demagnetizing field of the N{\'{e}}el wall):
\begin{equation}v_{W} = \gamma_{0} \frac{\Delta}{\alpha} H_{W}\approx\frac{\pi}{2} \gamma_{0} \Delta H_{DMI} = 
\frac{\pi}{2} \gamma\frac{D}{M_{s}},
    \end{equation}
where $\gamma$ is the gyromagnetic ratio, $\gamma_0 = \mu_0 \gamma$ and $H_{DMI} = D/(\mu_0 M_\mathrm{S}\Delta)$ is 
the DMI field that stabilizes the DW in the N{\'{e}}el configuration. 
They also agree with the results of domain wall velocity measurements \citep{Pham2016}.
\begin{figure*}[t]
\includegraphics[width=1\textwidth]{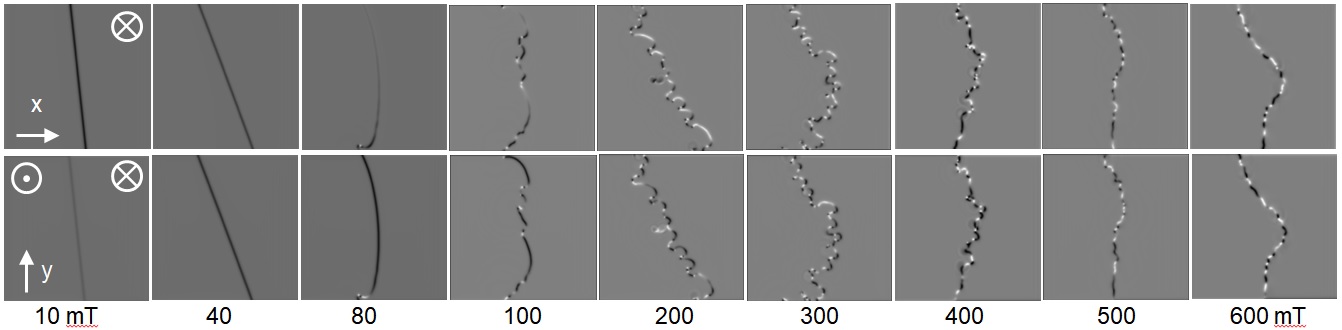}
\caption{\label{fig:DWsnapshots}
Sequence of images of the normalized magnetization components $m_x$ and $m_y$
(along and perpendicular to the strip direction) in $(1 \times 1)~\mu\text{m}^2$
moving windows for different driving fields.
The orientation of the magnetization vector in both domains is indicated in the first frame.
The displayed magnetization component is indicated by an arrow (white= parallel, black= antiparallel to arrow).
Each  frame is captured more than 50 ns after field application, when the DW has reached a stable configuration.
Simulation parameters are: $M_\text{s}$ = 0.756~MA/m,  $D$ = 1.5 mJ/m$^{2}$,
$K_\text{u}$ = 0.80~MJ/m$^{3}$, $A_\text{ex}$ = 16~pJ/m.
}
\end{figure*}

The simulations also show that, in agreement with experiments, the velocity is nearly constant for a certain
field range above the Walker field, and that the length (in field scale) of this velocity plateau is also
dependent on  $D$ and $M_\text{s}$.
In the case of strong DMI [e.g. $D=1.5$~mJ/m$^{2}$ as in Fig.~\ref{fig:speed-Ms-DMI}(b)], one sees that the
plateau length decreases as $M_\text{s}$  increases.
On the other hand, Fig.~\ref{fig:speed-Ms-DMI}(a) shows that for a constant $M_\text{s}$=1.01~MA/m, the length
of the plateau decreases as $D$ decreases, the plateau even disappearing for low DMI values
(e.g. $D\leq{0.5}$~mJ/m$^{2}$).
Therefore, a first parameter determining both the saturation speed and the field extension of the velocity plateau
appears to be the ratio between the DMI strength and the spontaneous magnetization.
Experimentally, both the saturation velocities and the plateau extension indeed decrease as this ratio
decreases (see Fig.~\ref{fig:Fig1-speeds} and Tab.~\ref{tab:param}).
Figure~\ref{fig:speed-Ms-DMI} in addition shows that the plateau length is also related to the domain wall 
width $\pi\sqrt{A_\text{ex}/K_\text{eff}}$.
To understand the microscopic mechanism leading to the velocity plateau and its breakdown, the DWs need
to be examined more closely by following the evolution of their magnetic structure and shape in different
applied fields and as a function of time.

Figure~\ref{fig:DWsnapshots} shows, for increasing magnitude of the applied field $B_z$, images of the
magnetization components in the $x$ and $y$ directions, in the case of a strip with $M_\text{s}=0.756$~MA/m,
$K_\text{u}= 0.80$~MJ/m$^3$ so that $K_\text{eff}= 0.44$~MJ/m$^{3}$ and $D = 1.5$~mJ/m$^2$.
Several trends can be identified, corresponding to the different dynamical regimes shown by the
corresponding  velocity curve [Fig.~\ref{fig:speed-Ms-DMI}(b), middle curve].
For low fields and up to the Walker field ($\simeq 77$~mT), the domain wall is straight and tilted with
respect to the stripe direction, the tilt angle initially increasing with field \citep{Boulle2013}.
Also, the DW magnetization rotates from the (left-handed) N{\'{e}}el wall orientation and becomes close to a
(right-handed) Bloch wall orientation as the Walker field is approached, as expected \cite{Thiaville2012}
(see image at 80~mT).
Above the Walker field, a sudden transition to a turbulent regime occurs.
The domain wall length increases with respect to the low field case and acquires a meander shape, with an
increasing number of curls, up to the end of the velocity plateau at around 350~mT.
This behaviour has already been observed in Refs.~\onlinecite{Yoshimura2015,Yamada2015,Pham2016}.
After reaching the maximum length, at the end of the plateau, the DW rapidly shortens but remains meandering.
With increasing field, it slowly straightens and its magnetization reaches a coherent precessional regime at
around 0.5~T.
In parallel, the in-plane magnetization images reveal a multiplication of sign changes.
When the DW is straight, this directly means that the DW magnetization spatially rotates between the Bloch
and N{\'{e}}el orientations.
On the other hand, when the DW is strongly meandering, the local magnetization orientation, i.e. its angle
$\varphi$ relative to the DW tangent, should be scrutinized in order to determine if the DW is N{\'{e}}el or
Bloch, and if a VBL is present.
This analysis is detailed in the Appendix.

\begin{figure*}[t]
\includegraphics[width=0.8\textwidth]{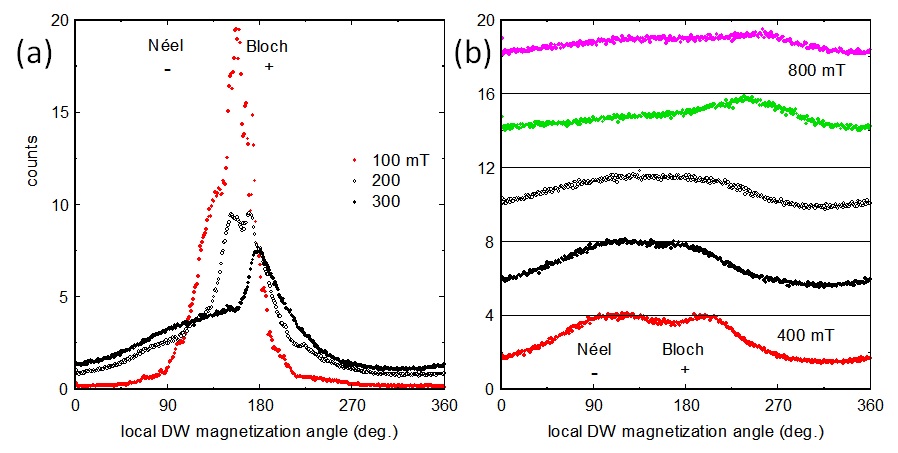}
\caption{\label{fig:histos}
Distribution of local DW magnetization angle $\varphi$ for several applied fields, within the velocity
plateau (a) and above it (b).
Parameters are $M_\text{s} = 0.756$~MA/m, $K_\text{eff}$ = 0.44 MJ/m$^{3}$ and $D = 1.5~\text{mJ/m}^2$.
For each field, the $10 \times 50$ acquired DW snapshots were processed, realizing a space- and
time-average of the DW magnetization angle.
The histogram binning is 1 degree, and the vertical scale is for 1000 counts in total.
In (b), curves are vertically offset (by 4 units) for clarity.
}
\end{figure*}

In order to have a global view of the DW magnetic structure, the histograms of local DW magnetization angles are
shown in Fig.~\ref{fig:histos}, separately for the plateau region (80-350~mT) and beyond the plateau.
Just above the Walker field, the distribution has a strong peak close to $\varphi=\pi$, the right-handed Bloch
wall derived from the left-handed N{\'{e}}el wall by $\pi/2$ precession.
It is therefore justified to describe the DW as a chiral Bloch wall, as proposed previously
\cite{Yoshimura2015,Yamada2015}.
As field increases, the intensity of this peak diminishes, whereas the background intensity raises, with a
preference for the energetically-favored left-handed N{\'{e}}el wall.
On the other hand, at large fields the histogram is increasingly flat.

\begin{figure}
\includegraphics[width=1\columnwidth]{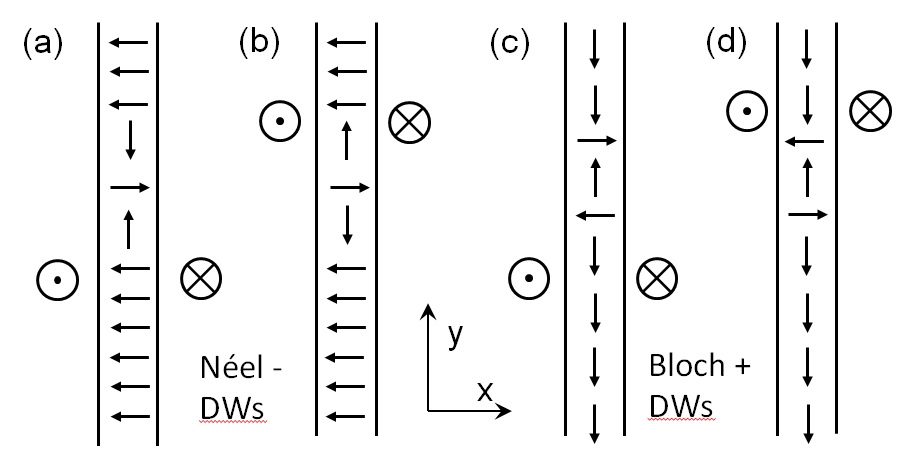}
\caption{\label{fig:schemaVBL}
Scheme of $2 \pi$ VBLs in Dzyaloshinskii domain walls, in statics (a,b) and around
the Walker field under positive $z$ field drive (c,d).
The winding of the $2 \pi$ VBL is opposite between (a) and (b), and (c) and (d).
}
\end{figure}
From the profiles of DW magnetization local angle $\varphi(s)$, the processes
involving the vertical Bloch lines were investigated.
As known from bubble physics \cite{Malozemoff1979}, in achiral Bloch walls 4 types of VBLs exist, degenerate
in energy and grouped into two senses of winding \cite{note-VBL}, corresponding to $\pm \pi$ rotations of $\varphi$,
the stable orientations for the DW moment being $\varphi=0, \pi$.
In the presence of interfacial DMI, it was proposed \cite{Yoshimura2015} that these 4 VBL types separate into
those favored, with a core magnetization parallel to the DMI effective field, and those that are unfavored.
This scheme implicitely rests on the concept of an underlying achiral Bloch wall.
This is however not the static state, which due to interfacial DMI is a chiral N{\'{e}}el wall.
Dynamically, just before the Walker breakdown, this DW does transform into a Bloch wall, but
it is chiral as it derives from the rest structure by $\pi/2$ precession around the applied field
\cite{Thiaville2012,Yoshimura2015}.
Hence, when DMI dominates over DW internal magnetostatics, only a $2\pi$ VBL is a relevant object. 
In static conditions it separates two chiral N{\'{e}}el walls with $\pm 2 \pi$ difference in $\varphi$ 
[Figs.~\ref{fig:schemaVBL}(a,b)], while just below the Walker threshold it separates two
chiral Bloch walls [Figs.~\ref{fig:schemaVBL}(c,d)] (this conclusion was also reached by Ref.~\onlinecite{Cheng2018}).
Note that, in the latter situation, if one fictively decomposes the $2 \pi$ VBL into two 1$\pi$ VBLs, a slower
magnetization rotation (larger width) is indeed observed for the VBL favored by DMI compared to the unfavored one
[Fig.~\ref{fig:profs}], in accordance with the arguments of Ref.~\onlinecite{Yoshimura2015}.
Therefore, we used as operational $2 \pi$ VBL definition the fact that $\varphi$ changes by $2 \pi$, the sense of
variation giving the VBL winding.
This procedure, starting from the $y$-bottom of the DW where the DW magnetization also varies with time,
cannot be used to locate precisely the VBLs, but at least is accurate for determining the number of VBLs and their winding.

\begin{figure}
\includegraphics[width=1\columnwidth]{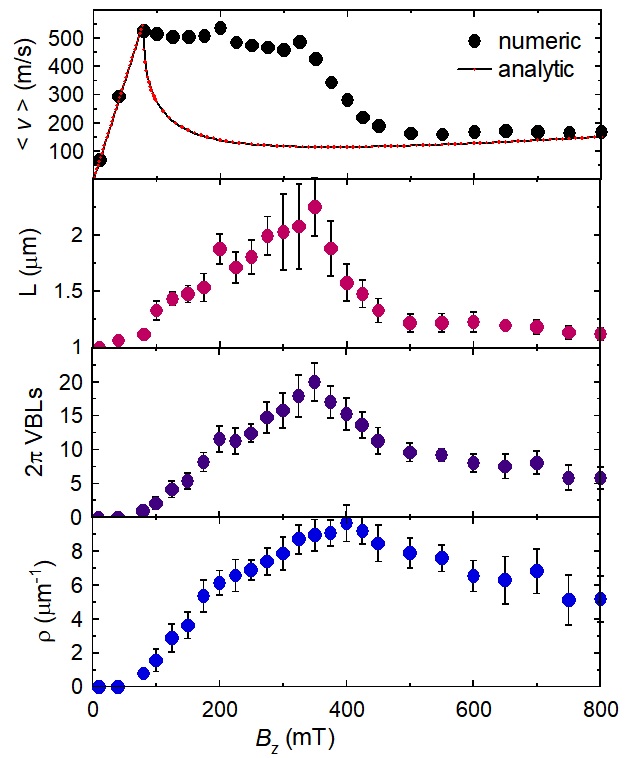}
\caption{\label{fig:solitons}
Field dependency of DW parameters: velocity, length, number of $2 \pi$ VBLs inside the
DW, and $2 \pi$ VBL density $\rho$ in 1~$\mu$m wide strip, with
$M_\text{s} = 0.756$~MA/m, $K_\text{eff}$ = 0.44 MJ/m$^{3}$ and $D = 1.5~\text{mJ/m}^2$.
The red velocity curve expresses the analytical 1D model (in the limit where the DMI field is much larger than the
demagnetizing field of the N{\'{e}}el wall) for the corresponding micromagnetic parameters.
}
\end{figure}

Figure~{\ref{fig:solitons}} summarizes the results of the detailed analysis of the shape and magnetic profile
of the moving DWs.
It appears that the variation of the DW velocity is correlated to the DW length, to the total number of $2 \pi$ VBLs,
as well as, but less clearly, to the density $\rho$ of $2 \pi$ VBL pairs.
Below the Walker field, the domain wall is straight, and without VBLs.
Above the Walker field but within the velocity plateau, the domain wall length and the number of $2 \pi$ VBLs increase.
Both decrease abruptly, like the velocity, when the plateau ends.
On the other hand, the $2 \pi$ VBL density increases inside the plateau but peaks after it has terminated, and then
slowly decreases.
After the plateau, the velocity progressively merges into the precessional velocity regime with low mobility,
as predicted by the one-dimensional ($q$, $\Phi$) model.

In order to see if this $2 \pi$ VBL density is a relevant parameter, we compare the maximum $2 \pi$ VBL pair densities
accommodated by the DW to the estimated $2 \pi$ VBL size.
An approximate calculation of its extent can be made under the assumptions that the domain wall remains straight, and
that the DMI dominates the domain wall internal demagnetizing effect.
The resulting energy takes the same form as that of the classical Bloch wall profile, only with an angle scaled by a factor 2.
We thus obtain that the $2 \pi$ VBL width $S$ (using the conventional Lilley definition in term of the tangent to the
angle profile at the central point \cite{Hubert1998}) reads
\begin{equation}
S=2 \pi \sqrt{{A_\text{ex} \Delta} / {(\pi D)}}.
\label{eq:s}
\end{equation}
Compared to these analytical models, full numerical calculations show that the domain wall bends at the VBL pair for
magnetostatic reasons, a feature noticed long ago \cite{Hubert1974b,Nakatani1987}, and analytically obtained recently
\cite{Cheng2018}.
One gets that the maximum $2 \pi$ VBL density is about a quarter of the inverse size.
This means that the end of the velocity plateau is not given by the uniform rotation model \cite{Malozemoff1979}
in which VBLs are densely packed.

\section{Discussion}
\label{sec:discu}

\begin{figure}[t]
\includegraphics[width=0.6\columnwidth]{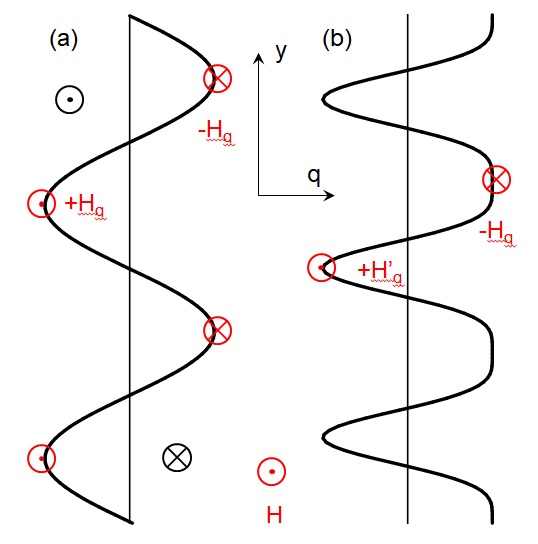}
\caption{
\label{fig:schemaDW}
Sketch of corrugated domain walls.
For a sinusoidal profile (a), the curvature-induced fields [Eq.~(\ref{eq:Hq})] are equal and opposite at
the two extremities of the DW, and a larger total field is applied to the part of the DW lagging behind.
For a corrugated profile with smaller zones lagging behind (b), these experience a yet larger field.
This last configuration resembles the images of Fig.~\ref{fig:DWsnapshots}.
Magnetizations are drawn in black and fields in red.}
\end{figure}

A framework to analyze the DW dynamics above the Walker field is afforded by the Slonczewski
equations \cite{Slonczewski1972,Slonczewski1973}.
For the present situation, their variables are the local DW $x$ position $q(y,t)$, a function of $y$ position
across the strip and time $t$, and the angle of DW magnetization $\Phi(y,t)$, in the absolute frame.
For large corrugations of the DW, one should rather use $q(s,t)$ and $\Phi(s,t)$ where $s$ is the curvilinear abscissa
and the DW local displacement $q$ is measured along the normal to the DW \cite{Slonczewski1974,Malozemoff1979}, but we
forget this subtlety in this qualitative analysis.
When the DW is not straight, the equations indicate that the DW surface energy induces an additional field $H_q$
parallel to the applied field, proportional to the DW curvature (in magnitude and sign)
\begin{equation}
\label{eq:Hq}
H_q = \frac{\sigma}{2 \mu_0 M_\mathrm{S}} \frac{\partial^2 q}{\partial y^2},
\end{equation}
where $\sigma=4\sqrt{A_\mathrm{ex} K_\mathrm{eff}}-\pi D$ is the DW surface energy density.
Note that this field also appears in the creep theory of magnetic domain walls \cite{Lemerle1998}.
As shown in Fig.~\ref{fig:schemaDW}, $H_q$ decreases the drive field at the places that are ahead of the average DW
position, and increases it at the places that lag behind.
The typical value of this field is not at all negligible: for the parameters of Fig.~\ref{fig:solitons} one has
$\sigma= 5.9$~~mJ/m$^2$, so that a sine modulation of amplitude $\pm 50$~nm and wavelength 200~nm produces
$\mu_0 H_q = 192$~mT.

Based on these ideas, a heuristic `corrugated wall' regime was proposed by J. Slonczewski \cite{Slonczewski1972}
for the DW dynamics above the Walker field, in the negative mobility region,
in which $H + H_q \leq H_\mathrm{W}$ for the DW parts which are ahead, whereas in the lagging parts
$H + H_q \gg H_\mathrm{W}$.
The increased field in the lagging parts causes a faster precession of DW magnetization, hence drives the
nucleation of VBLs, corresponding to the observations.
As a corrugated DW is longer, the increase of both DW length and VBL number seen in Fig.~\ref{fig:solitons}
is consistent with this mechanism.
The corrugated wall picture also corresponds well to the DW images shown in Fig.~\ref{fig:DWsnapshots}, for
$\mu_0 H_z= 100$ and 200~mT.
In Ref.~\onlinecite{Slonczewski1972} Slonczewski then looked for a steady-state DW corrugation and associated
velocity.
The numerical simulations show, however, that the situation is more complex so that this model cannot be
directly applied.
Nevertheless, despite the complexity of the DW shape and magnetization profile, some integral relations hold
that help understanding this regime.

The first is the `momentum conservation' i.e. the spatial and temporal
average of the first Slonczewski equation, given by
\begin{equation}
\label{eq:momentum}
<<\frac{\partial \Phi}{\partial t}>> + \frac{\alpha}{\Delta} <<\frac{\partial q}{\partial t}>>
= \gamma_0 H,
\end{equation}
where the double brackets indicate the two averages.
This is the version of the DW dynamics model under the assumption of a constant DW width parameter $\Delta$, but
it has been shown that this equation can be generalized to arbitrary spin textures \cite{Thiaville2007},
upon definition of a generalized $\Phi$ angle and DW position.
In the present case, if $H > H_\mathrm{W}$ and the DW moves at an average velocity $v \approx v_\mathrm{W}$, one gets
$<<\partial \Phi / \partial t>> \approx \gamma_0 \left( H - H_\mathrm{W} \right)$.
This means that DW magnetization precession is required for fast DW motion above the Walker field, contrarily
to a naive expectation that DW magnetization precession equals velocity breakdown.
In the corrugated DW model, this precession is localized in the lagging DW parts, and the momentum conservation
is satisfied if the length fraction $f$ of these parts and the curvature-induced field $H_q$ satisfy
$f \sqrt{\left( H + H_q \right) ^2 - H_\mathrm{W}^2} \approx H - H_\mathrm{W}$, a relation that can be approximated to
$\left( H + H_q \right) f \approx H - H_\mathrm{W}$ for large H.
Inspection of Fig.~\ref{fig:DWsnapshots}, for $\mu_0 H_z= 100$ and 200~mT reveals that $f$
indeed increases with $H$.

The second relation is simply the energy balance \cite{Slonczewski1972}: by the DW motion the total energy of
the sample decreases, at a rate $- 2 \mu_0 M_\mathrm{S} H v$ per unit cross-section of the strip.
In the micromagnetic dynamics equation, energy decreases only by the damping term, with a volumic
dissipative power given by
$-\alpha \left( \mu_0 M_\mathrm{S} / \gamma_0 \right) \left( \partial \vec{m} / \partial t \right)^2$.
The obvious source of magnetization time variation is the DW, and the two DW mobility regimes can be
obtained by evaluating this energy dissipation.
But, in addition, the DW can emit spin waves that disappear inside the domains, dissipating energy also.
As detailed in the Appendix, spin waves are emitted when a $2 \pi$ VBL disappears,
releasing its energy.
At rest, still neglecting deformation of the DW profile in its vicinity, the energy of a $2 \pi$ VBL
is analytically evaluated as
\begin{equation}
\label{eq:E-VBL}
\lambda = 16 \sqrt{A_\mathrm{ex} \Delta \pi D }
\end{equation}
per unit length (the thickness of the film).
The number of $2 \pi$ VBLs disappearing per unit time can be estimated from the time evolution of the profiles 
of the local angle $\varphi$ (see Appendix).
We find that the energy dissipated by this process amounts to about one third of the required additional energy
dissipation compared to that of a 1D DW moving at the Walker velocity, namely
$- 2 \mu_0 M_\mathrm{S} \left( H - H_\mathrm{W} \right) v_\mathrm{W}$.
In fact, the energy dissipated by the destruction of the $2 \pi$ VBLs is at least twice as large as this estimation, as
the comparison of numerical and analytical $\varphi(s)$ profiles shows that the dynamical VBLs are compressed
by a factor larger than 4, leading to a larger dissipation upon annihilation.
This shows that the VBL-based energy loss mechanism is quantitatively dominant in the velocity plateau
at the Walker velocity.

We finally discuss the end of this plateau.
A first remark is that the field where the plateau ends is close to the field where the 1D model predicts
the velocity to be minimum (see Fig.~\ref{fig:solitons}).
This field was considered apparently only by J.~C. Slonczewski \cite{Slonczewski1972,Slonczewski1972b}, and
may therefore be called the Slonczewski field $H_\mathrm{S}$.
In the 1D model with purely second-degree DW effective anisotropy $K_{DW}$ leading to a DW internal anisotropy
field $H_{K_{DW}}=2K_{DW}/(\mu_0 M_\mathrm{S})$, which is analytically solvable, one has
$H_\mathrm{W} = \alpha H_{K_{DW}}/2$ and $v_\mathrm{W} = \gamma_0 \Delta H_{K_{DW}} /2$ for the Walker threshold, whereas
for the Slonczewski minimum the quantities are \cite{Slonczewski1972}
\begin{eqnarray}
H_\mathrm{S} &=& H_{K_{DW}}\frac{1+\alpha^2}{2 \sqrt{2+\alpha^2}} \approx H_{K_{DW}}/(2 \sqrt{2}); \\
v_\mathrm{S} &=& \gamma_0 \Delta H_{K_{DW}} \frac{\alpha \sqrt{2+\alpha^2}}{2(1+\alpha^2)}
\approx \gamma_0 \Delta H_{K_{DW}} \frac{\alpha}{2},
\end{eqnarray}
the approximations holding for $\alpha \ll 1$.

In our case, when the DMI-induced field satisfies $H_D \gg H_{K_{DW}}$ this 1D model applies (just consider
$\Phi/2$ as the angular variable), and $H_\mathrm{W}\approx \alpha H_D$ \cite{Thiaville2012}, so that one gets
\begin{equation}
\label{eq:H_S_DDW}
H_\mathrm{S}=\frac{1+\alpha^2}{\sqrt{2+\alpha^2}} H_D =
\frac{\pi(1+\alpha^2)}{2 \sqrt{2+\alpha^2}} \frac{D}{\mu_0 M_\mathrm{S} \Delta},
\end{equation}
where the numerical factor is $1.11$ in the small $\alpha$ limit, so that $H_\mathrm{S}\approx ~ H_\text{DMI}$
(remember that $H_D = (\pi / 2) H_\text{DMI}$).

One sees that the dependence of $H_\mathrm{S}$ on the micromagnetic parameters $D$ and $M_{s}$ is exactly that 
found in the experiments and the simulations, and moreover that the agreement with both is quantitatively very good. 
It also confirms that the length of the velocity plateau is related to the DW width, as already found in the simulations.
This can be taken as a hint that the corrugated wall model is relevant for the dynamics above the Walker field,
the plateau lasting at most up to the field where the driving force for the corrugation instability disappears.

Regarding the value of the plateau velocity, we only have qualitative arguments.
Within the corrugated wall model, the corrugation instability cannot grow beyond reaching $H+H_q=H_\mathrm{W}$ in
the leading edge parts, so that one has $v \le v_\mathrm{W}$.
We have also seen that the momentum and energy conservation relations can be satisfied in this model, by localized precession
in the lagging behind parts and $2 \pi$ VBL destruction, respectively, when the DW moves close to the Walker velocity.
This shows that $v \approx v_\mathrm{W}$ is an admissible solution in the negative mobility regime.
The specificity of large DMI values is that, on top of a large Walker field and Walker velocity,
the DMI forces the formation of $2 \pi$ VBLs, which can store a large amount of energy which can be released upon 
annihilation.
A large DMI also leads to a large Slonczewski field, the apparent end of the velocity plateau, which therefore becomes 
easier to observe.

\section{\label{sec:conclusion} Conclusion}

We have experimentally shown that, for chiral N{\'{e}}el walls stabilized by the interfacial
Dzyaloshinskii-Moriya interaction (Dzyaloshinskii domain walls), the maximum velocity and the end
of the high velocity plateau are mainly controlled by the ratio $D/M_\text{s}$ of DMI to
spontaneous magnetization.
This observation has been confirmed by systematic 2D micromagnetic simulations.
These additionally revealed that the domain wall width also impacts the end of the high velocity plateau,
and that the moving domain wall is strongly corrugated in the plateau region.
A detailed analysis of the domain wall in-plane magnetization variation along its length has shown the
key role of $2 \pi$ vertical Bloch lines, textures that are topologically stable and therefore
disappear through a Bloch point, dissipating  a large energy through the emission of spin waves.
We propose that this plateau corresponds to the negative mobility regime of the one-dimensional
domain wall dynamics, as qualitatively described by the corrugated domain wall model
of J.~Slonczewski.
As a result, the end of the high velocity plateau is simply proportional to the effective DMI field, with a numerical factor independent of damping in the low-damping limit.

\begin{acknowledgments}
This work was realized with the support of the Erasmus+ programme of the European Union.
S.~P., J.~V. and A.~T. acknowledge the support of the Agence Nationale de la Recherche, projects
ANR-14-CE26-0012 (ULTRASKY) and ANR-17-CE24-0025 (TOPSKY).
We thank B. Fernandez, Ph. David and E. Mossang for technical help.
D.S.C was supported by a CNPq Scholarship (Brazil).
A.~T. thanks J. Miltat, S. Rohart and J. Sampaio for inspiring discussions.

\end{acknowledgments}

\section{Appendix: analysis of the magnetic structure of the domain walls}
\label{sec:App}

\begin{figure*}[t]
\includegraphics[width=1\textwidth]{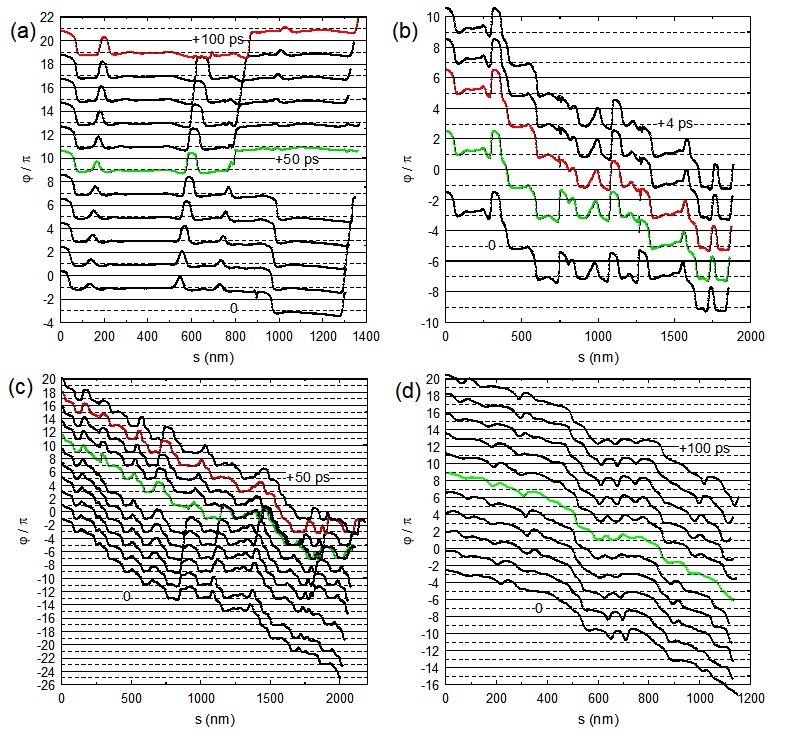}
\caption{\label{fig:profs}
Time evolution of the profiles of local magnetization angle $\varphi$ for the DWs presented in Fig.~\ref{fig:DWsnapshots}.
The applied fields are (a) $B_z=100$~mT (a), 200~mT (b), 300~mT (c) and 500~mT (d).
The relative times (ps) are indicated, and the profiles are successively offset by (multiples of) $2 \pi$, for clarity.
The dashed horizontal lines indicate the dynamically-favored chiral Bloch wall orientations.
Profiles are colored to highlight changes or help differentiating them.
The origin of the curvilinear abscissa ($s$) is at the bottom of the images.}
\end{figure*}

The profiles of the local magnetization angle $\varphi$ for the DWs presented in Fig.~\ref{fig:DWsnapshots} at 
zero relative time, together with their time evolution, are plotted in Fig.~\ref{fig:profs}, for selected field values.
The local DW magnetization angles are counted relative to the DW tangent direction (globally along $+y$
as the DW is followed from bottom to top), in the trigonometric sense so that $\varphi= \pi/2$ is the
left-handed N{\'{e}}el wall and $\varphi= \pi$ the right-handed Bloch wall.
One notices that for all profiles $\varphi$ globally decreases with increasing curvilinear abscissa $s$.
Taking into account that, in the presence of DMI, the DW magnetization precession starts at a given edge of
the strip \cite{Yamada2015} (here the bottom edge, see image at 80~mT in Fig.~\ref{fig:DWsnapshots}), and
that the applied field is positive, this can be rationalized.

In Fig.~\ref{fig:profs}(a), corresponding to $B_z=100$~mT, the progressive precession of DW magnetization
is shown: a precession from $\simeq 0.3 \pi$ to $\simeq 1.5 \pi$ takes place at $s \approx 150$~nm,
and a precession from $\simeq 1 \pi$ to $2 \pi$ is observed at $s \approx 600$~nm.
In the latter case, the two $2 \pi$ VBLs that have formed disappear (at 1~ps interval) between the last two
profiles.
The last curve shows a signature of the spin-wave wake emitted after the $2 \pi$ VBLs destruction.
This corresponds to the main mode observed.
More rarely, and especially just above the Walker field, the conversion of one VBL of the pair into another
one, with thus opposite winding and core magnetization, is seen.
These two lines, having opposite winding senses, merge back into a $2 \pi$ VBL, which eventually disappears
as shown above.
Such process is observed at $s \approx 800$~nm.

Figure~\ref{fig:profs}(b), corresponding to $B_z=200$~mT, shows that the $2 \pi$ VBL destruction
is extremely fast.
In fact, the process is instantaneous as the annihilation of the topologically stable $2 \pi$ VBL
involves a Bloch point crossing the sample thickness \cite{Malozemoff1979,Thiaville2018}.
As the sample is described by one layer of cells, this Bloch point is numerically virtual (there
is no interval between layers of mesh points where it could sit).
One $2 \pi$ VBL disappears between times 30 and 31~ps, at $s \approx 1250$~nm, and another one between times
31 and 32~ps, at $s \approx 750$~nm.
The propagation of the spin-wave wake is also observed on the $\varphi(s)$ profiles.

For a larger field $B_z= 300$~mT [Fig.~\ref{fig:profs}(c)], still within the velocity plateau, the same
processes are seen.
The VBLs are however less visible as the overall slope of $\varphi(s)$ has increased.
Finally, for  $B_z= 500$~mT [Fig.~\ref{fig:profs}(d)], beyond the velocity plateau, the dominant time
evolution is the global precession of the DW magnetization, but $2 \pi$ VBL destructions occur from time
to time.
We thus have a global picture of the time evolution of the angles $\varphi$: they increase with time, in a
non-uniform way by the formation of unwinding VBL pairs, the continued precession transforming
these pairs into $2 \times 2 \pi$ `pairs of pairs'.
Each $2 \pi$ line can disappear through the passage of a Bloch point, releasing its energy by spin waves
emission.

\end{document}